\documentclass[reprint,amsmath,amssymb,aps,showkeys]{revtex4-2}

\usepackage{amsmath,amssymb}
\usepackage{graphicx}
\usepackage{xcolor}
\usepackage{hyperref}

\usepackage{upgreek}
\usepackage{mathrsfs} 
\DeclareFontFamily{OT1}{pzc}{}
\DeclareFontShape{OT1}{pzc}{m}{it}{<-> s * [1.1] pzcmi7t}{}
\DeclareMathAlphabet{\mathpzc}{OT1}{pzc}{m}{it}

\newcommand{\Diff}{\mathrm{Diff}}

\newcommand{\bs}{\boldsymbol}
\newcommand{\id}{\mathrm{id}}
\newcommand{\diff}{\mathrm{diff}}

\newcommand{\hh}{\mathcal{H}}
\newcommand{\G}{\mathfrak{G}}
\newcommand{\vphi}{\varphi}
\newcommand{\rarrow}{\rightarrow}
\renewcommand\-{^{-1}}

\begin{document}

\title{Invariant Path-Integral Quantization and Anomaly Cancellation} 

\author{J. François} \email{jordan.francois@uni-graz.at} \affiliation{University of Graz (Uni Graz), Heinrichstraße 26/5, 8010 Graz, Austria, and \\ Masaryk University (MUNI), Kotlářská 267/2, Veveří, Brno, Czech Republic, and \\ Mons University (UMONS), 20 Place du Parc, 7000 Mons, Belgium. } 
\author{L. Ravera} \email{lucrezia.ravera@polito.it} \affiliation{Politecnico di Torino (PoliTo), C.so Duca degli Abruzzi 24, 10129 Torino, Italy, and \\ Istituto Nazionale di Fisica Nucleare (INFN), Section of Torino, Via P. Giuria 1, 10125 Torino, Italy, and \\ Grupo de Investigación en Física Teórica (GIFT), Universidad Cat\'{o}lica De La Sant\'{i}sima Concepci\'{o}n, Alonso de Ribera 2850, Concepción, Chile. } 

\date{\today}

\begin{abstract}
We present an invariant relational path-integral quantization framework for general-relativistic gauge field theories based on the Dressing Field Method. 
The construction implements an automatic anomaly-cancellation mechanism that encompasses Bardeen-Wess-Zumino counterterms. 
The resulting framework unifies invariant schemes across contexts ranging from electroweak theory to cosmology, and is amenable to lattice implementations, key to high-precision tests in both domains.
\end{abstract}


\maketitle

\section{Introduction}
\label{Introduction}

The identification and quantization of the physical degrees of freedom (d.o.f.) is a foundational issue of gauge field theory. 
Standard approaches based on gauge fixing, such as the BRST formalism, provide powerful quantization schemes but are hindered by well-known limitations, including Gribov-Singer obstructions \cite{Gribov,Singer1978, Singer1981} and the introduction of auxiliary ghost sectors, which can obscure the physical content of the theory.

An alternative strategy is to construct physical, gauge-invariant d.o.f. directly, to be then subjected to quantization. 
The Dressing Field Method (DFM)  \cite{GaugeInvCompFields,JTF-Ravera2024gRGFT} is a wide ranging,  systematic and geometrically transparent framework to achieve this. 
It is particularly well suited to general-relativistic gauge field theory (gRGFT), where it provides variables invariant under both gauge and diffeomorphism transformations. 
It is  background independent, intrinsically non-perturbative, yet admits a perturbative implementation, and naturally enjoys a relational interpretation \cite{JTF-Ravera-Foundation2025}.  
The DFM is intrinsically formulated in field space, which provides a modern geometric framework for classical and quantum field theory \cite{DeWitt2003}. 

In this letter, we present an  \emph{invariant path-integral quantization} of gRGFTs based on the DFM. 
The framework bypasses gauge fixing entirely.
As we shall discuss, it systematizes and unifies several recent  developments in the gRGFT literature. 

We first formulate gRGFT in field space, introducing its \emph{cocyclic geometry}, in which anomalies admit a natural characterization. 
After introducing the basics of the DFM, we present the dressed path-integral quantization. 

The latter features an \emph{intrinsic anomaly-cancellation mechanism}, which generates generalized Bardeen-Wess-Zumino counterterms \cite{Bertlmann, Bonora2023}. 
Whenever anomalies encode physically significant information, it must be preserved in the invariant formulation. This occurs through an ``anomaly seesaw mechanism", whereby anomalies of the covariant (bare) formulation give way to anomalies associated with transformations of the dressing fields, capturing \emph{physical reference frame changes}.

Another central result is the derivation of the dressed BRST algebra satisfied by dressed variables, with its \emph{dressed ghost}, controlling potential residual symmetries.
It trivializes upon complete symmetry reduction via the DFM. 
This definitely clarifies the issue of the relation between dressing and the BRST framework (raised, e.g., by \cite{Grassi:2024vkb, Freidel:2026stu}).
We conclude by showcasing the scope and versatility of the framework, through broad applications ranging from electroweak and Higgs physics to cosmological perturbation theory. 

These results establish a clear and timely advance toward a novel geometric framework for the fully invariant and physically transparent quantization of gRGFTs.

\section{Field space and cocycles}
\label{Field space and cocycles}

The field space $\Phi$ of a gRGFT is an infinite-dimensional manifold whose points are  collections of fields on an $m$-dimensional manifold $M$, noted $\upphi=\{A, \vphi, e/g, \ldots\}$ and comprising gauge potentials $A$, matter fields $\vphi$,  a metric $g$ or tetrad $e$, etc.
It supports the right action of $\G:=\Diff(M) \ltimes \hh$,  where $\Diff(M)$ is the group of diffeomorphisms of $M$ and 
\mbox{$\hh := \{ \gamma: M\rarrow H\, |\, {\gamma}^{\!\gamma'}={\gamma'}\- \gamma \gamma'\}$} is the ``internal" gauge group, with $H$ a (finite-dimensional) Lie group. 
The semi-direct product of $\G$ is  
$(\psi, \gamma)\cdot(\psi', \gamma')= \big(\psi \circ \psi', \, \gamma(\psi^{-1*} \gamma') \big)$.
The  
action is 
$\upphi\mapsto
\psi^*(\upphi^\gamma)$, where the pullback action by $\psi \in \Diff(M)$ is clear, while $\upphi^\gamma$ is the action by $\hh$. 
%\footnote{It is, e.g., for a gauge potential $A^\gamma=\gamma\-A\gamma+\gamma\- d\gamma$, and for a matter field $\varphi^\gamma =\gamma\-\varphi$. If we keep track of the representation $\rho$ of $H$ in which the matter field is valued, we write $\varphi^\gamma =\rho(\gamma)\-\varphi$. If $\hh=\mathcal{SO}(1,3)$ is the Lorentz group, the tetrad 1-form $e$ taking values in the fundamental representation, we have  that $e^\gamma =\gamma\-e$.}.

This action (being free in  realistic physical situations) makes the field space $\Phi$ a principal fiber bundle with structure group $\G$,  whose fibers are the $\G$-orbits of fields $\upphi$,  with canonical projection  $\pi: \Phi \rightarrow  \Phi/\G=:\mathcal M$.
The  moduli space $\mathcal M$ \emph{abstractly} representing  \emph{physical d.o.f.}

The action of the Lie algebra Lie$\G=\diff(M) \oplus \text{Lie}\hh$ on $\Phi$ is $\upphi \mapsto \delta_{(X, \lambda)} \upphi :=  \big(\mathfrak L_X \upphi, \delta_\lambda \upphi \big)$, 
with $\mathfrak L_X$ the Lie derivative along the vector field $X \in \diff(M)$.

The \emph{gauge group} of $\Phi$ is   $\bs\G$, whose elements are functionals $(\bs\psi, \bs\gamma):\Phi \rarrow \G$  \mbox{transforming  under $\G$ as}:
$(\bs\psi, \bs\gamma)^{(\psi, \gamma)}
:=({\bs\psi}[\upphi^{(\psi, \gamma)}], {\bs\gamma}[\upphi^{(\psi, \gamma)}])
=\text{Conj}_{(\psi, \gamma)\-}(\bs\psi, \bs\gamma)$, 
with $\text{Conj}$ the conjugation action of $\G$ on itself. 
%\footnote{The inverse of  $(\psi, \gamma)$ being $(\psi, \gamma)\- = (\psi\-, \psi^* \gamma\-)$.}.
In the  literature,  elements of $\bs \G$ are often called ``field-dependent" local transformations,  $(\bs\psi, \bs\gamma)=({\bs\psi}[\upphi], {\bs\gamma}[\upphi])$.
Elements of $\G$ are thus said to be ``field-independent". 
Correspondingly, $(\bs X, \bs \lambda)
\in$ Lie$\bs\G$.
The gauge group $\bs\G$ is the full covariance group of gRGFTs -- as foresaw first, to the best of our knowledge, by Bergmann and Komar \cite{Bergmann-Komar1972} in the case of General Relativity (GR), when considering metric-dependent diffeomorphisms.

The field space $\Phi$ has 
a space of differential forms $\Omega^\bullet(\Phi)$; we denote $\bs\alpha$ a $k$-form.
The action of $\G$ and Lie$\bs\G$  on forms, noted $\bs\alpha^{(\psi, \gamma)}$ and $\delta_{(X, \lambda)} \bs\alpha$, 
defines their \emph{equivariance}. 
Similarly, the action of $\bs \G$ and Lie$\bs\G$ on forms,  $\bs\alpha^{(\bs\psi, \bs\gamma)}$ and 
$\delta_{(\bs X, \bs\lambda)} \bs\alpha$, define their \emph{gauge transformation}.
Two types of forms are especially  relevant to physics.

First, \emph{cocyclic} tensorial forms $\Omega_\text{tens}^\bullet(\Phi, C)$ are those whose gauge transformation is
\begin{align}
\label{cocycl-GT}
\bs\alpha^{(\bs\psi, \bs\gamma)}
=
C\big(\upphi ; (\bs \psi, \bs\gamma)\big)\- \bs \alpha,
\end{align}
featuring a \emph{1-cocycle} on $\G$, i.e., a map $C:\Phi \times \G \rarrow \mathbb{C}$ defined by the property
\begin{align*}
C\big(\upphi ; (\psi, \gamma)\cdot(\psi', \gamma') \big) =  C\big(\upphi ; (\psi, \gamma)\big)\, C\big(\upphi^{(\psi, \gamma)} ; (\psi', \gamma')  \big), 
\end{align*}
consistent with the right action of $\G$ 
\footnote{Remark that  1-cocycles  generalize representations: a $\upphi$-independent cocycle $C$ being just a group morphism.
%This means that cocyclic tensorial forms generalize \emph{standard} tensorial forms, transforming via representations (of the structure group).
}.
We may use the compact notation $C(\bs\psi, \bs\gamma) = C\big(\upphi ; (\bs \psi[\upphi], \bs\gamma[\upphi])\big)$.
Infinitesimally, \eqref{cocycl-GT} yields 
$\delta_{(\bs X, \bs\lambda)} \bs\alpha
= - a\big((\bs X, \bs\lambda) ; \upphi\big)\, \bs\alpha$,
with a Lie$\G$ 1-cocycle 
$a: \Phi \times \text{Lie}\G \rarrow \mathbb C$, 
whose defining property is the linear version of that of  $C$. 

The second key space is that of \emph{basic} forms $\Omega_\text{basic}^\bullet(\Phi)$, 
which are those that are strictly $\bs\G$-invariant, $\bs\alpha^{(\bs\psi, \bs\gamma)}
= \alpha$, or $\delta_{(\bs X, \bs\lambda)} \bs\alpha=0$.
%\begin{align}
%\bs\alpha^{(\bs\psi, \bs\gamma)}
%= \alpha, \quad \text{or}
%\quad
%\delta_{(\bs X, \bs\lambda)} \bs\alpha=0.
%\end{align}
They  \emph{concretely} represent  physical d.o.f. 
%\footnote{Notice that a form that is \emph{invariant}, $\bs\alpha \in \Omega^\bullet_\text{inv}(\Phi)$, i.e., with trivial $\G$-equivariance $R^\star_{(\psi, \gamma)} \bs\alpha=\bs\alpha$, is not automatically $\boldsymbol{\G}$-invariant, and thus cannot qualify as a physical observable (basic form). For a form to be basic, it has to be both invariant and \emph{horizontal}, $\bs\alpha \in \Omega^\bullet_\text{hor}(\Phi)$, i.e., such that $\iota_{\mathcal X^v}\bs\alpha=0$, $\forall \, \mathcal X^v$ vector tangent to the orbit.}

\smallskip
When $\bs \alpha$ is both a $k$-form on $\Phi$ and a $p$-form of $M$, it transforms under $\Diff(M)$ as $\bs\alpha^\psi= \psi^* \bs\alpha$. 
It can be integrated on $p$-dimensional regions $U\subset M$ to yield a $k$-form on $\Phi$: $\langle \bs\alpha, U \rangle:= \int_U \bs \alpha$. 
Regions of $M$ provide the ``defining representation" of $\Diff(M)$, which acts (on the right) as $U \mapsto \psi\-(U)=:U^\psi$. 
Integrals are by definition $\Diff(M)$-invariant:
$\langle \psi^*\bs\alpha, \psi\-(U) \rangle=\langle \bs\alpha, U \rangle$. 
But regions $U$ are of course $\hh$-singlets, $U^\gamma=U$, so integrals are not automatically $\hh$-invariant: 
$\langle \bs\alpha^\gamma, U^\gamma \rangle\neq\langle \bs\alpha^\gamma, U \rangle$.
They are only if $\bs\alpha$ is $\hh$ invariant/basic.
Remark that it means that the integral of a $\G$-basic $\bs\alpha$ cannot be basic itself: 
one needs a \emph{basic integration theory}.

\subsection{Quantum dynamics}
\label{Quantum dynamics}

The above geometric structure gives the kinematics of gRGFTs. 
Their dynamics is given by a  functional (i.e., $0$-forms) on $\Phi$:  by a Lagrangian or action  classically, then  by a path integral quantum mechanically. 

A Lagrangian $m$-form is $L:\Phi \rarrow \Omega^m(M)$, $\upphi \mapsto L(\upphi)$, with associated action $S:\Phi \rarrow \mathbb R$, $\upphi \mapsto S(\upphi):=\int_U L(\upphi)$, where $U \subset M$ is a compact region. 
Let us define the functional $Z_\text{cl}:\Phi \rarrow \mathbb C$, 
$\upphi \mapsto Z_\text{cl}(\upphi):= \exp{\tfrac{i}{\hbar} S(\phi)}$.
Then, the quantum theory is specified by a path integral $Z[\upphi]:=\int \mathcal D\upphi\, Z_\text{cl} =:\exp\tfrac{i}{\hbar} W[\upphi]$, with $\mathcal D\upphi$ a formal integration measure on  $\Phi$, and  
where
$W[\upphi]:= -i\hbar\ln  Z[\upphi]$ is the connected  functional. 
The vacuum expectation  value (v.e.v.) of a  (polynomial) functional $\mathcal F[\upphi]$ is defined by
$\langle \mathcal F \rangle:= Z[\upphi]\- \int \mathcal D\upphi\, \mathcal F[\upphi] Z_\text{cl}$.

Introducing auxiliary currents $J$ ``sourcing" fields $\upphi$, and defining 
$Z[\upphi, J]:=\int \mathcal D\upphi\, 
\exp\{\tfrac{i}{\hbar} \int J \, \upphi \} Z_\text{cl} =:\exp\tfrac{i}{\hbar} W[\upphi, J]$,
one may  compute the v.e.v. of $F[\upphi]$ as: 
$\langle \mathcal F \rangle:=Z[\upphi]\- \mathcal F[-i\hbar \tfrac{\delta}{\delta J}]\, Z[\upphi, J]\big|_{J=0}$. 
The \emph{connected} v.e.v. is:
$\langle  \mathcal F \rangle_c:= \mathcal F[-i\hbar \tfrac{\delta}{\delta J}]\, W[\upphi, J]\big|_{J=0}$. 
In particular, (connected) $n$-point correlation functions  $\langle \upphi_1 \ldots \upphi_n\rangle_{(c)}$ are obtained for $F[\upphi]=\upphi_1 \ldots \upphi_n$, where $\upphi_i=\upphi(x_i)$ and $x_i \in M$: 1-point functions are the v.e.v. of fields $\langle \upphi\rangle$, 2-point functions give propagators $\langle \upphi_1  \upphi_2\rangle_{(c)}$, etc.

Starting with a classical theory  $Z_\text{cl} \in \Omega^0_\text{basic}(\Phi)$,  
and if the measure $\mathcal{D}\upphi$ is basic,  we have  $Z \in \Omega^0_\text{basic}(\Phi)$ for the quantum theory.
Meaning we have  \mbox{$\delta_{(\bs X, \bs\lambda)} Z=0$}, and $\delta_{(\bs X, \bs\lambda)} W=0$  expresses a ``generalized" Ward-Takahashi (WT) \mbox{identity} -- the standard one is $\delta_{(X, \lambda)} W=0$, for $\G$.

If the measure is not  basic, the  theory is \emph{anomalous}; $Z \in \Omega^0_\text{tens}(\Phi, C)$, i.e., $Z$ has $\bs\G$-transformations 
\begin{equation}
\label{anomalous-Z-W}
Z^{(\bs\psi, \bs\gamma)}
=C(\bs\psi, \bs\gamma)\- Z,
\end{equation}
with $C=\exp\{-\tfrac{i}{\hbar} c\}$
a $\G$ 1-cocycle; 
the connected functional  $\bs\G$-transforms as 
$W^{(\bs\psi, \bs\gamma)}=W + c(\bs\psi, \bs\gamma)$.
Linearly, 
\begin{equation}
\label{Linear-anom-Z-W}
\delta_{(\bs X, \bs\lambda)} Z
= - a(\bs X, \bs\lambda) Z, 
\, \text{ or }\, 
\delta_{(\bs X, \bs\lambda)} W
=  i\hbar\, a(\bs X, \bs\lambda), 
\end{equation}
where $a(\bs X, \bs\lambda):=-\tfrac{i}{\hbar}\tfrac{d}{d\tau}\, c (\bs\psi_\tau, \bs\gamma_\tau)|_{\tau=0}$ is the \emph{quantum anomaly}, whose Lie$\G$ 1-cocycle defining property is known as the Wess-Zumino consistency condition.
Eq. \eqref{Linear-anom-Z-W} expresses the generalized anomalous WT identity.
\smallskip

\paragraph{Gauge fixing.}

Even with a basic measure $\mathcal D\upphi$, $Z$ integrates over the volume of $\G$, making it divergent.
A technical patch is to gauge-fix: 
Geometrically, this means selecting a local section $\sigma: \mathcal U  \rarrow \Phi_{|\mathcal U}$, over a domain $\mathcal U \subset \mathcal M$, cutting orbits once, and satisfying by definition $\pi \circ \sigma=\id_\mathcal{U}$.
Its image is  Im$(\sigma) :=\{\upphi\in \Phi\, |\, \mathcal C(\upphi)=0\}$ for some constraints $\mathcal C(\upphi)=0$. 
Gauge-fixing a functional, a 0-forms $\mathcal F\in \Omega^0(\Phi)$, is restricting it on Im$(\sigma)$, $\mathcal F^\text{gf}:=\mathcal F_{|\text{Im}(\sigma)}$, which amounts to 
a pullback by $\sigma$: $\mathcal F^\text{gf}\simeq \sigma^\star \mathcal F= \mathcal F \circ \sigma \in \Omega^0(\mathcal U)$.
%\footnote{This geometric view shows transparently how, conceptually, gauge fixing is in some sense equivalent to working with physical d.o.f., on $\mathcal U\subset \mathcal M$. The Gribov-Singer obstruction \cite{Gribov, Singer1978, Singer1981} is the fact that there exists no  global section $\sigma:\mathcal M  \rarrow \Phi$ (i.e., no `good' gauge fixing) when $\Phi$ is a non-trivial bundle: gauge fixing is only possible `locally' on limited domains $\mathcal U \subset \mathcal M$.}.
A gauge-fixed quantum theory is $Z^\text{gf}:=\sigma^\star Z =  Z \circ \sigma$. 
BRST gauge fixing achieves this through a subtle variant of Lagrange multipliers.
\smallskip

\paragraph{BRST framework.}

For later purpose, we review the BRST formalism.
The nilpotent BRST operator $s$, such that $s^2\equiv 0$, reproduces the action of $\delta_{(X, \lambda)}$ on $\upphi$ with $(X, \lambda)$ replaced by Lie$\G$-valued ghosts $(\xi, v)$. Both $s$ and $(\xi, v)$ are given ghost degree ($gh$) 1, and $0$ for $d$ and $\upphi$, so that $\big(\upphi, (\xi, v)\big)$ equipped with $d+s$, such that $(d+s)^2\equiv0$, form a bigraded differential algebra in form degree on $M$ and $gh$ degree, adding as  total degree $deg$. 

The BRST transformations of the fields $\upphi$ and ghosts are $s\,\upphi =
\big(s_\text{diff} \upphi, s_\text{gauge} \upphi \big)
= \big(\mathfrak L_\xi \upphi, (-)^{deg \upphi}\delta_v \upphi \big)$
and  \mbox{$s(\xi, v)=
-\tfrac{1}{2}\big[(\xi, v),(\xi, v)\big]_{\text{Lie}\G}
= \big(-\tfrac{1}{2}\mathfrak L_\xi \xi, \  \mathfrak L_\xi v - \tfrac{1}{2}[v, v] \big)$}
$= \big(-\tfrac{1}{2}[\xi, \xi], \  \xi(v) - \tfrac{1}{2}[v, v] \big)
$, respectively, reproducing the semi-direct structure of Lie$\G$ and ensuring $s^2=0$.
The BRST complex $\big(\Lambda^{gh}, s \big)$ is essentially the Chevalley-Eilenberg (CE) complex of Lie$\G$,  with coefficient in functionals of $\upphi$, i.e., in $\Omega^0(\Phi)$ \cite{Bonora-Pasti-Bregola, Bonora-Cotta-Ramusino-Reina,DeAzc-Izq}, so that the BRST ($\sim$CE) cohomology  $H^{gh}(s)$ captures non-trivial $\G$-invariants. 
In particular, anomalies $a\big( (\xi, v); \upphi \big) \in H^1 (s)$, and $sa=0$ is just the defining 1-cocycle property of $a$ (i.e., the Wess-Zumino consistency condition).
While $H^0(s)\simeq \Omega^0_\text{basic}(\Phi)$, basic forms captures invariants of higher form degrees on $\Phi$.

\section{Relational quantization via  DFM}
\label{Relational quantization via  DFM}

From what precedes, it is clear that key problems are defused if one  operates  with  basic forms.
A basic path integral would be finite, so one  avoids gauge fixing.  
It would also be automatically free of $\G$-anomalies \footnote{As is clear from the fact that the BRST cohomology is `orthogonal' to basic cohomology.}.
A~basic path integral would  define what  may be called \emph{basic quantization}, or $\bs\G$-\emph{invariant quantization}. 
We here achieve it via the Dressing Field Method (DFM),  which is a systematic way to realize basic forms on $\Phi$.
It  does so, furthermore, in a manifestly \emph{relational} way \cite{JTF-Ravera2024gRGFT,JTF-Ravera-Foundation2025}.

A key notion of the method is that of $\G$-\emph{dressing field}, a couple of maps $(\upsilon, u)$ with $\upsilon:N \rarrow M$, where $N$ is a model manifold (often $N=\mathbb R^m$), and $u: M \rarrow H$, by definition $\G$-transforming as 
\begin{align}
\label{dressing-def}
\upsilon^{(\psi, \gamma)}:=\psi\- \circ \upsilon
\ \text{ and }\ 
u^{(\psi, \gamma)}:=\psi^*(\gamma\- u), %\notag\\[.5mm]
%&\text{or } \quad (\upsilon,  u)^{(\psi, \gamma)}:=(\psi, \gamma)\-\cdot( \upsilon,  u).
\end{align}
or $(\upsilon,  u)^{(\psi, \gamma)}:=
(\psi, \gamma)\-\cdot
( \upsilon,  u)$.
Remark that this definition
means  $(\upsilon, u)\notin \G$. 
Let us denote $\mathcal Dr$ the space of dressing fields.
Then, a \emph{field-dependent dressing field} is a functional 
$(\bs\upsilon, \bs u): \Phi \rarrow \mathcal Dr$, $\upphi \mapsto \big(\bs\upsilon(\upphi), \bs u(\upphi) \big)$, with $\G$-equivariance given by \eqref{dressing-def}, thus with $\bs\G$-transformation
\begin{align}
\label{dressing-def-2}
(\bs \upsilon, \bs u)^{(\bs\psi, \bs\gamma)}:=
(\bs\psi, \gamma)\-\cdot
(\bs \upsilon, \bs u).
\end{align}
Correspondingly, 
$\delta_{(\bs X, \bs \lambda)}(\bs \upsilon, \bs u)
=
\big( -\bs X \circ \bs \upsilon, \ \, \mathfrak L_{\bs X} \bs u\, -\bs \lambda \bs u  \big)$. 
Notice again that \eqref{dressing-def-2}
implies  $(\bs\upsilon, \bs u)\notin \bs\G$.
A $\upphi$-dependent $\bs\G$-dressing field  induces the \emph{dressing map}
\begin{equation}
\label{Dressing-map}
\begin{aligned}
F_{(\bs \upsilon, \bs u)}: \Phi &\rarrow \Phi^{(\bs \upsilon, \bs u)}\simeq \mathcal M, \\
\upphi &\mapsto \upphi^{(\bs \upsilon, \bs u)}:= {\bs\upsilon}^* (\upphi^{\bs u})=:\bar\upphi.
\end{aligned}
\end{equation}
It is a coordinatization of the  moduli base space $\mathcal M$ via composite, \emph{dressed fields} $\upphi^{(\bs \upsilon, \bs u)}={\bs\upsilon}^* (\upphi^{\bs u})=\bar\upphi$, which are $\G$ and $\bs\G$ invariant.

Since $(\bs \upsilon, \bs u)$ is $\upphi$-dependent, the dressed fields $\upphi^{(\bs \upsilon, \bs u)}$ have a natural relational interpretation: some d.o.f. within the bare fields $\upphi$, entering $(\bs \upsilon, \bs u)$, are used as a \emph{physical reference frame} to coordinatize the others. 
Dressed fields are  general invariant relational variables,  
the dressing map realizes a relational coordinatization. 
Furthermore, 
 since $(\bs\upsilon, \bs u)\notin \bs\G$, dressed fields 
are \emph{not} \mbox{elements} of $\Phi$, in particular they 
are not gauge fixings of bare fields $\upphi$. 
Gauge fixing and dressing are technically and conceptually distinct operations \cite{Berghofer-Francois2024}.
\smallskip

The DFM ``rule of thumb" is the following.
To find the basic, dressed, counterpart of a form $\bs\alpha$ on $\Phi$, first determine its $\bs\G$-gauge transformation $\bs\alpha^{(\bs \psi, \bs \gamma)}$
(\emph{not} merely its $\G$-equivariance), 
then substitute in the resulting expression  the element $(\bs \psi, \bs \gamma)\in \bs\G$ by the dressing field $(\bs \upsilon, \bs u)$: 
one then gets $\bs\alpha^{(\bs \upsilon, \bs u)} \in \Omega^\bullet_\text{basic}(\Phi)$.
The latter's \emph{functional} expression in terms of invariant dressed fields $\bar\upphi=\upphi^{(\bs \upsilon, \bs u)}$ is identical to that of the bare counterpart $\bs\alpha$ in terms of bare fields $\upphi$.
In particular, if $\bs\alpha \in \Omega^\bullet_\text{tens}(\Phi, C)$, then its dressed version is 
\begin{align}
\label{Dressed-cocyclic-tens-form}
\bs\alpha^{(\bs \upsilon, \bs u)}= C(\bs\upsilon; \bs u)\- \bs\alpha\  \in \Omega^\bullet_\text{basic}(\Phi),
\end{align}
with $C(\bs\upsilon, \bs u)$ a \emph{cocyclic dressing field}, which transforms under $\G$, or  Lie$\G$, by definition as
\begin{equation}
\label{cocyclic-dressing-field}
\begin{aligned}
C(\bs\upsilon, \bs u)^{(\psi, \gamma)}&=
C(\psi, \gamma)\-
C(\bs\upsilon, \bs u),\\[1mm]
\delta_{(\bs X, \bs\lambda)} C(\bs\upsilon, \bs u)
&= - a( X, \lambda) C(\bs\upsilon, \bs u).
\end{aligned}
\end{equation}

The rule of thumb works likewise for integrals $\langle \bs \alpha, U\rangle$, whereby one gets \emph{dressed integrals}
\begin{align}
\label{dressed-int}
\langle \bs \alpha, U\rangle ^{(\bs \upsilon, \bs u)} :=
\langle\bs\alpha^{(\bs \upsilon, \bs u)}, U^{\bs\upsilon} \rangle,
\end{align}
$U^{\bs\upsilon}:=\bs \upsilon\-(U)$ being $\Diff(M)$-invariant \mbox{\emph{dressed regions}}, namely \emph{relationally} defined $\upphi$-dependent regions of the \emph{physical} spacetime $M^{\bs\upsilon}$ \footnote{More precisely, $M^{\bs\upsilon}$ is the physical manifold of events, the physical spacetime being  $(M^{\bs\upsilon}, g^{\bs\upsilon})$, where $g^{\bs\upsilon}$ is the $\Diff(M)$-invariant physical metric field.} (implementing  Einstein's point-coincidence argument), where the  dressed fields $\bar\upphi=\upphi^{(\bs \upsilon, \bs u)}$ ``live".
Notice that \emph{each entry} in the pairing \eqref{dressed-int} is $\bs\G$-invariant,  yielding the \mbox{\emph{basic integration theory}}  alluded to earlier, which thus means  integration over physical spacetime.

\subsection{Invariant path-integral quantization}

Relational quantization is  defined by the \emph{dressed path integral} (PI) 
\begin{align}
\label{Dressed-PI}
Z[\upphi]^{(\bs \upsilon, \bs u)}=Z[\bar \upphi]=\int \mathcal D\bar \upphi \, Z_\text{cl}(\bar\upphi)=\exp \tfrac{i}{\hbar}W[\bar \upphi],
\end{align}
where $Z_\text{cl}(\bar\upphi)=Z_\text{cl}(\upphi)^{(\bs \upsilon, \bs u)} = \exp \tfrac{i}{\hbar} S[\bar\upphi] \in \Omega^0_\text{basic}(\Phi)$, with
$S[\bar\upphi]=S(\upphi)^{(\bs \upsilon, \bs u)}= \int_{U^{\bs\upsilon}} 
L(\bar\upphi)=\int_{U^{\bs\upsilon}}L(\upphi)^{(\bs \upsilon, \bs u)}$, is the invariant, dressed, relational classical theory.
The v.e.v. of a dressed polynomial functional 
$\bar{\mathcal F}:=\mathcal F[\bar \upphi]$ is then
$\langle \bar{\mathcal F} \rangle:= Z[\bar \upphi]\- \int \mathcal D\bar \upphi\, \mathcal F[\bar \upphi] Z_\text{cl}(\bar\upphi)$.

The latter can be computed via the basic generating functional 
\begin{equation}
\begin{aligned}
Z[\bar \upphi, J^{\bs \upsilon}]:=&\int \mathcal D\bar \upphi\, 
\exp\Big\lbrace{\tfrac{i}{\hbar} \int_{U^{\bs\upsilon}} J^{\bs\upsilon} \,  \bar \upphi \Big\rbrace} Z_\text{cl}(\bar\upphi) \\
=&:\,\exp\tfrac{i}{\hbar} W[\bar \upphi, J^{\bs \upsilon}],
\end{aligned}
\end{equation}
with dressed currents \mbox{$J^{\bs\upsilon}:=\bs\upsilon^* J$}, $\Diff(M)$ (thus $\G$) invariant auxiliary variables with respect to which one performs variational differentiation to get the dressed v.e.v.:
$\langle \bar{\mathcal F} \rangle:=Z[\bar \upphi]\- \mathcal F[-i\hbar \tfrac{\delta}{\delta J^{\bs \upsilon}}]\, Z[\bar \upphi, J^{\bs\upsilon}]\big|_{J^{\bs\upsilon}=0}$.
Idem for the  connected v.e.v.:
$\langle \bar{\mathcal F} \rangle_c:= \mathcal F[-i\hbar \tfrac{\delta}{\delta J^{\bs \upsilon}}]\, W[\bar \upphi, J^{\bs\upsilon}]\big|_{J^{\bs\upsilon}=0}$. 

Dressed (connected) $n$-point correlation functions $\langle \upphi_1 \ldots \upphi_n\rangle_{(c)}$ are then found for   $F[\bar\upphi]=\bar \upphi_1 \ldots \bar \upphi_n$, 
with $\bar \upphi_i=\bar \upphi (\bar x_i)$, and where $\bar x_i:=\bs\upsilon\-(x_i) \in M^{\bs\upsilon}$ are $\Diff(M)$-invariant, relationally defined physical spacetime events.
The v.e.v. of dressed fields is then $\langle \bar \upphi\rangle=\langle \upphi^{(\bs \upsilon, \bs u)}\rangle$, while dressed propagators are  $\langle \bar \upphi_1  \bar \upphi_2\rangle_{(c)}$, and similarly for higher dressed correlation functions.

Given a metric field $g$ on $M$, the dressed $\Diff(M)$-invariant physical metric $g^{\bs\upsilon}$ gives a  relational causal structure,  defining causal and chronological partial orders between events of the physical spacetime $M^{\bs\upsilon}$. 
Then, the dressed correlation functions for non-gravitational fields  ($\bar \upphi \neq g^{\bs\upsilon}$) define an invariant normal (``time") ordering $\langle \bar \upphi_1 \ldots \bar\upphi_n\rangle_{(c)}=\langle T\{\bar \upphi_1 \ldots \bar\upphi_n \}\rangle_{(c)}$ in the standard way.
\smallskip

\paragraph{Anomaly cancellation.}

The dressed PI $Z^{(\bs\upsilon, \bs u)}$ is finite, the volume of $\G$ being factored out; \mbox{gauge fixing} is unnecessary (impossible, even).
Furthermore, since a dressing operation and a $\G$-transformation are \emph{formally} functionally alike, if the bare measure $\mathcal D\upphi$ is $\G$-invariant it is equal to the dressed measure, i.e., to the measure on dressed fields: $\mathcal D\bar \upphi=\mathcal D\upphi^{(\bs\upsilon, \bs u)}=\mathcal D\upphi$.

In case $\mathcal D\upphi$ is anomalous, i.e., if $Z \in \Omega^0_\text{tens}(\Phi, C)$ and satisfies \eqref{anomalous-Z-W}, 
the dressed PI is a  case of \eqref{Dressed-cocyclic-tens-form}, 
\begin{equation}
\label{Dressed-PI-2}
\begin{aligned}
Z^{(\bs \upsilon, \bs u)}&= C(\bs\upsilon; \bs u)\- Z \  \in \Omega^\bullet_\text{basic}(\Phi),\\
\exp\{\tfrac{i}{\hbar}W^{(\bs\upsilon, \bs u)} \}
&= 
\exp{\tfrac{i}{\hbar}\{W + c(\bs\upsilon, \bs u)\}},
\end{aligned}
\end{equation}
with the cocyclic dressing field $C(\bs\upsilon; \bs u)$ satisfying \eqref{cocyclic-dressing-field}, \begin{equation}
\label{cocyclic-dressing-kills-anomaly}
\begin{aligned}
C(\bs\upsilon, \bs u)^{(\bs\psi, \bs\gamma)}&=
C(\bs\psi, \bs\gamma)\-
C(\bs\upsilon, \bs u),\\
\text{ or }\, 
\delta_{(\bs X, \bs\lambda)}\, c(\bs\upsilon, \bs u)
&= - i\hbar\,a(\bs X, \bs\lambda).
\end{aligned}
\end{equation}
That is, the cocyclic dressing exactly compensates the quantum anomaly of the bare anomalous $Z$, so that the dressed, relational PI $Z^{(\bs \upsilon, \bs u)}$ automatically implements an \emph{anomaly-cancellation mechanism}. 

When the dressing field is \emph{ad hoc}, i.e., $\upphi$-independent -- so $(\upsilon, u)$ are introduced by hand as extra fields -- then $C(\upsilon, u)=\exp -\tfrac{i}{\hbar}c(\upsilon, u)$ reproduces exactly the Wess-Zumino counterterms often introduced to eliminate anomalies \footnote{Such \emph{ad hoc} dressing fields $(\upsilon, u)$ can be understood as generalized Stueckelberg fields;  though the latter are usually introduced to \emph{implement} (artificially) a local symmetry, not to eliminate one, as in the DFM. See \cite{JTF-Ravera2024gRGFT}.} -- see, e.g., section 12.3 of \cite{GockSchuck}, chap. 15 of \cite{Bonora2023}, or  chap. 4 of \cite{Bertlmann}.

The DFM achieves anomaly cancellation from a first-principle and relational approach. 
This can be further seen by considering how the DFM yields a \emph{dressed BRST algebra} that trivializes when $\G$-dressing fields are used.
\smallskip

\paragraph{Dressed BRST formalism.}

We shall simply state the result -- detailed in a companion paper \cite{JTF-Ravera2026-RQcompanion} --
that dressed fields $\bar \upphi=\upphi^{(\bs\upsilon, \bs u)}$ satisfy a dressed BRST algebra
\begin{equation}
\label{Dressed-BRST}
\begin{aligned} 
s\bar \upphi 
&=
\big(s_\text{diff} \bar \upphi, s_\text{gauge} \bar \upphi \big)
= \big(\mathfrak L_{\bar\xi} \bar\upphi, (-)^{deg \bar \upphi}\delta_{\bar v} \bar\upphi \big),  \\[1mm] 
s(\bar\xi, \bar v)
&=
-\tfrac{1}{2}\big[(\bar\xi, \bar v),(\bar \xi, \bar v)\big], 
\end{aligned}
\end{equation}
where the bracket has the same semi-direct structure as that of {Lie}$\G$, and where the \emph{dressed ghosts} are defined as
\begin{align}
(\bar \xi, \bar v) 
&=
\big(\xi^{(\bs\upsilon, \bs u)}, v^{(\bs\upsilon, \bs u)} \big) \\
&= (\bs\upsilon, \bs u)\- \cdot (\xi,  v)\cdot (\bs\upsilon, \bs u)
+  (\bs\upsilon, \bs u)\- \cdot s(\bs\upsilon, \bs u) \notag\\
=&
\left( 
\bs\upsilon\-_* \xi \circ \bs  \upsilon + \bs\upsilon\-_* s_{\text{diff}} \bs\upsilon, \ \,
\upsilon^*(\bs u \- v \bs u + \bs u \- s_\text{gauge} \bs u)
\right). \notag
\end{align}
The dressed BRST algebra encodes residual symmetries remaining when  only a subgroup $\bs{\mathfrak{J}}\subset\bs\G$ is reduced via a $\bs{\mathfrak{J}}$-dressing field. 
But when a $\bs\G$-dressing field is used, since its defining property is, in BRST language, 
\begin{align}
\label{BRST-dressing-field}
s (\bs\upsilon, \bs u) 
&= \big(s_{\text{diff}} \bs\upsilon,\  s_{\text{diff}}\bs u + s_\text{gauge} \bs u \big) \notag\\
&= \big( -\xi \circ  \bs\upsilon,\ \mathfrak{L}_{\xi}\bs u -v \bs u \big),
\end{align}
the dressed ghost \emph{vanishes  identically}, $(\bar\xi, \bar v)=0$. 
So, the  dressed BRST algebra \eqref{Dressed-BRST} trivializes, reflecting, indeed, the $\G$-invariance of dressed fields $\bar\upphi$.
We thus have that the cohomology for the dressed BRST algebra is trivial, which is another way to express the fact that the dressed quantum theory $Z^{(\bs \upsilon, \bs u)}$ is anomaly free.  
The triviality of the dressed BRST algebra also shows (again) that BRST gauge fixing is not only unnecessary, but impossible. 
 
This answers the question raised by \cite{Freidel:2026stu} regarding the links between BRST formalism (or BRST gauge fixing) and dressing. 
This also settles the conjecture of \cite{Grassi:2024vkb} as to dressing being an ``intertwiner" trivializing the BRST operator (or charge): 
The dressing map \eqref{Dressing-map} is a relational (if the dressing  $\upphi$-dependent, not \emph{ad hoc}) instantiation of the projection of field space $\Phi$, $F_{(\bs\upsilon, \bs u)} \sim \pi$, realizing  basic form $\Omega^\bullet_\text{basic}(\Phi)$ cohomology, which is ``orthogonal" to BRST cohomology (coinciding only in their degree 0), whereby $\G$-anomalies are ``killed".

Yet, $\bs \G$-anomalies may carry non-trivial  information,  their cancellation  imposing relevant physical constraints. 
Any such physical information is not lost in this framework, but recurs in the form of anomalies for what in the DFM are called \emph{transformations of the second kind},  parametrizing possible ``ambiguity" in the choice of dressing field, and encoding \emph{physical frame covariance}. 
This topic is discussed in Appendix \ref{Frame Covariance and Anomaly Seesaw Mechanism}.

\section{Discussion}
\label{Discussion}

This general relational framework unifies a number of invariant schemes in the literature, old and new,  which it reproduces as special cases: e.g., Stueckelberg fields \cite{Ruegg-Ruiz}
and edge modes à la \cite{DonnellyFreidel2016} (which are instances of \emph{ad hoc} dressing fields), dressing time \cite{Freidel:2026stu}, gravitational dressings à la \cite{Giddings-et-al2006, Giddings-Perkins2024}, dynamical reference frames à la \cite{Carrozza-Hoehn2021}, etc.
We highlight here only three particularly relevant cases. 
\smallskip

\paragraph{4D Green-Schwarz mechanism and axion.}

First, we notice that \cite{Alvarez-Gaume-Vazquez-Mozo2023} sketches what is presented as illustrating the Green-Schwarz (GS) mechanism for the theory of an $\hh=\mathcal{U}(1)$ gauge field $A$ coupled to chiral Weyl fermions $\uppsi$, whose anomaly is killed via  an `axion' field $\theta$. 
The latter is exactly an (Abelian) \emph{ad hoc} dressing field $\bs u=u=\exp i\theta$ (introduced alongside $\upphi=\{\uppsi, A\}$, not built from them), and the anomaly-canceling GS counter-term added to the action is precisely a cocyclic dressing  $c(\upphi; \bs u)=c(\upphi;\theta)$, satisfying the Abelian version of \eqref{cocyclic-dressing-field}.

This simple case indicates that  
the DFM extension to higher gauge theories (to be detailed in a forthcoming work), featuring the same  automatic cancellation of -- and seesaw mechanism for -- anomalies of higher form symmetries (e.g., in the context of SymTFT),  may encompass as a special case the GS mechanism of $10D$ heterotic or Type I supergravity/string theory: the 2-form $B$ would be an \emph{ad hoc} higher dressing field.
\smallskip

\paragraph{The electroweak model.}

Our framework reproduces invariant formulations of the electroweak (EW) model bypassing the notion of spontaneous symmetry breaking, in keeping with Elitzur theorem \cite{Elitzur1975}.
Notably the Fröhlich-Morchio-Strocchi (FMS) approach \cite{Frohlich-Morchio-Strocchi80}, which takes roots in  \cite{Hooft1980, Banks-Rabinovici1979}, as well as observations from Higgs \cite{Higgs66} and Kibble \cite{Kibble67}. 
In that case, the $\hh=\mathcal{SU}(2)$-dressing is built from the $\mathbb C^2$-scalar field $\varphi$, $\bs u=\bs u(\varphi)$, and used to dress the fields $\upphi=\{\varphi, (A, B), \uppsi, \ldots\}$ where $A$ and $B$ are a $U(1)$ and an $SU(2)$ gauge field, respectively, and $\uppsi$ is the chiral lepton doublet field; the dressed fields are the physical states $\bar\upphi=\{\varphi^{\bs u}, (A^{\bs u}, B^{\bs u}), \uppsi^u,\ldots \}=\{\upeta, (\gamma, W^\pm/Z^0), (e,\nu_e),\ldots\}$. 
Upon expansion around the minimum $\upeta_0=v$ of the potential $V(\upeta)$, $\upeta =v +H$, one identifies the physical scalar field $H$, and mass terms arise for the physical gauge fields $W^\pm/Z^0$ (through their minimal coupling to $\upeta$) and leptons $(e,\nu_e)$ (through Yukawa couplings with $\upeta$).

The invariant relational variables $\bar\upphi$ are essentially the FMS local operators \cite{Sondenheimer2020}, interpreted as ``bound states" of the ``elementary" (bare) fields $\upvarphi$ and $\{(A, B), \uppsi\}$ -- in analogy with QCD with quark/gluons forming color-singlet bound states.
Unfolding the 2-point functions of dressed fields $\langle \bar \upphi\bar \upphi\rangle_{(c)}$ in terms of the bare ones expanded around $\upeta =v +H$ reproduces the so-called FMS expansion of propagators  -- which is taken to explain how standard perturbation theory for gauge variant  fields $\upphi$ relates to the physical spectrum \cite{Maas2019}.

We further observe that EW theory on the lattice  uses tacitly  as key variables the dressed fields $\bar\upphi$ (see, e.g., \cite{Creutz2024, Creutz2025}), showing that our framework is a continuous formulation fitting naturally with Wilsonian lattice gauge theory and is readily amenable to lattice computations, which is key to high-precision test of Higgs physics. 
\smallskip

\paragraph{Cosmological perturbation theory.}

Specializing our approach to $\G=\Diff(M)$, we reproduce all manners of ``scalar coordinatization" in GR; either à la Komar-Bergmann \cite{Komar1958, Bergmann1961, Bergmann-Komar1960}, with a metric-dependent dressing field $\bs\upsilon=\bs\upsilon(g)$, or via material reference frames  as in \cite{DeWitt1960,Kuchar1980, Rovelli1991, Brown-Marolf1996}, with a matter/dust field $\phi$ dependent dressing, $\bs\upsilon=\bs\upsilon(\phi)$.

In the context of early cosmology, considering a metric $g=g_0+h$ with $g_0$ an homogeneous solution, and a matter density field $\phi=\phi_0+\rho$ with $\phi_0$ constant,  the dynamical fields are the perturbations $\upphi=\{h, \rho\}$.
If one extracts a $\Diff(M)$-dressing fields $\bs\upsilon=\bs\upsilon(h)$,  the dressed fields $\bar\upphi=\{h^{\bs\upsilon}, \rho^{\bs\upsilon}\}=\{\bar h, \bar\rho\}$ 
reproduce the Bardeen \cite{Bardeen1980} and Mukhanov-Sasaki variables \cite{Mukhanov1987, Sasaki1986}, which are the basis of an \emph{invariant cosmological perturbation theory}. A dressing $\bs\upsilon$ in that context has also been called a `clock' \cite{Giesel-et-al2018}.
The $2$-point correlation function of the relational dressed variables $\langle\bar\upphi\bar\upphi \rangle_{(c)}$  gives an invariant way to computate the power spectrum of the CMB. 
This is again a continuum invariant formulation amenable to lattice computation, useful for high-precision cosmology.

\begin{acknowledgments} 
J.F. is supported by the Austrian Science Fund (FWF), grant \mbox{[P 36542]}, 
and by the Czech Science Foundation (GAČR), grant GA24-10887S.
L.R. is supported by the project 
GrIFOS, YOUR MSCA SoE, 
CUP E13C24003600006, of which this work is part.
This study was supported by the European Union - NextGenerationEU, under the Italian National Recovery and Resilience Plan (PNRR), Mission 4 Component 2 Investment 1.2, funding scheme ``Young Researchers" (D.D. 201 del 3.7.2024). 
%This manuscript reflects only the authors’ views and opinions and the Ministry cannot be considered responsible for them.
\end{acknowledgments} 

%\section*{Author contribution declaration} 

%Both authors share equal credit for the work done in this paper: J.F. and L.R. conceived the idea presented, J.F. and L.R. developed the formalism and did the calculations. Both J.F. and L.R. authors contributed equally to the writing of the manuscript. 

%\section*{Funding declaration}

%The research was supported by the Austrian Science Fund (FWF), grant \mbox{P 36542}, the Czech Science Foundation (GAČR), grant GA24-10887S, and the GrIFOS research project, funded by MUR, Italy, CUP E13C24003600006, ID SOE2024$\_$0000103. 

\appendix

\section{Anomaly seesaw mechanism}
\label{Frame Covariance and Anomaly Seesaw Mechanism}

The DFM accounts systematically for \emph{a priori} ambiguities in the choice or construction of dressing fields \cite{JTF-Ravera2024gRGFT}.
Given their defining property \eqref{dressing-def}-\eqref{dressing-def-2}, two dressing fields have to be related by
\begin{align}
\label{gRGFT-ambig-dressing}
(\bs\upsilon', \,\bs u')
&= \big(\bs\upsilon \circ \upvarphi,\, \bs u\, (\bs\upsilon^{-1*} \bar\upzeta)\,  \big)  \notag\\
&=
(\bs\upsilon, \,\bs u) \cdot 
(\upvarphi, \,\bar\upzeta\,)
=:(\bs\upsilon, \bs u)^{(\upvarphi, \bar\upzeta)},
\end{align}
for $\upvarphi \in \Diff(N)$, with $N=$ Im$(\bs\upsilon^*)$ the source space of  $\bs\upsilon$, and $\upzeta \in \mathcal G:=\big\{ \upzeta: M \rarrow G\, |\, \upzeta^\gamma =\upzeta\big\}$ with $\gamma \in \hh$ and $H$ (the target space of $\bs u$), supporting the action of $G$ (e.g., $H\subseteq G$), so that $\bar\upzeta:= \bs\upsilon^* \upzeta \in \bar{\mathcal G} := \big\{\bar\upzeta: N \rarrow G\, |\, \bar\upzeta^{\bar\gamma} =\bar\upzeta\big\}$. 
This defines the action of the group $\bar\G:=\Diff(N)\ltimes \bar{\mathcal G}$ on the space of $\G$-dressing fields. 
The action of $\bar{\bs\G}$, $\upphi$-dependent elements of $\bar\G$, is defined similarly.
We shall call  $\bar{\bs\G}$ the group of \emph{transformation of the second kind}. 

The latter captures \emph{physical reference frame} changes.  
If dressings $\big(\bs \upsilon[\upphi_i], \bs u[\upphi_i]\big)$ and $\big(\bs \upsilon'[\upphi_j], \bs u'[\upphi_j]\big)$ are  built from distinct fields' d.o.f. $\upphi_i$ and $\upphi_j$ within  $\upphi$, 
one  \emph{defines} 
\begin{equation}
\label{ref-frame-cov}
(\bs\upvarphi, \,\bar{\bs\upzeta}\,)
:=
\big(\bs \upsilon[\upphi_i], \bs u[\upphi_i]\big)\- \cdot \big(\bs \upsilon'[\upphi_j], \bs u'[\upphi_j] \big)\ \in \bar{\bs\G}.
\end{equation}
Given the  relational view of the DFM, whereby  
$\upphi^{(\bs\upsilon, \bs u)}$ gives a relational description of  physical d.o.f. with respect to $\upphi_i$,  while $\upphi^{(\bs\upsilon', \bs u')}$ gives another with respect to $\upphi_j$,
\eqref{ref-frame-cov} is naturally interpreted  as  a \emph{change of physical reference frame}. 
Hence, $\bar{\bs\G}$-transformations  
implement the \emph{physical frame covariance} of the dressed formulation.

Regarding the action of $\bar\G$  (and $\bar{\bs\G})$ on dressed fields, two cases are to be distinguished.

$\bullet$ In the first case, $\bar\G$ leaves  the bare fields invariant,  $\upphi^{(\upvarphi,\, \bar\upzeta)}=\upphi$, 
so dressed fields $\bar\upphi=\upphi^{(\bs\upsilon,\, \bs u)}$ $\bar\G$-transform  as 
\begin{align*}
(\bar \upphi)^{(\upvarphi, \,\bar\upzeta)} 
:= 
\big(\upphi^{(\bs\upsilon,\, \bs u)}\big)^{(\upvarphi, \bar\upzeta)}
=
\big( \upphi^{(\upvarphi, \bar\upzeta)}\big)^{(\bs\upsilon, \bs u)^{(\upvarphi, \bar\upzeta)}} 
= \upvarphi^*\big({\bar\upphi}^{\bar\upzeta} \big),
\end{align*}
the $\bar{\mathcal G}$-transformations ${\bar\upphi}^{\bar\upzeta}$ of the dressed fields $\bar\upphi$ having the same \emph{functional} expressions as the $\hh$-transformations $\upphi^\gamma$ of the bare fields $\upphi$.
It follows that the cocyclic dressing $\bar\G$-transforms in a way mimicking \eqref{gRGFT-ambig-dressing}:
\begin{align}
\label{Cocyclic-dressing-2nd-kind}
C(\bs\upsilon,\, \bs u)^{(\upvarphi,\, \bar\upzeta\,)}
=C(\bs\upsilon,\, \bs u)\, C\big(\bar\upphi; (\upvarphi, \bar\upzeta\, )\big),
\end{align}
proven via $C$'s defining property. 
The term
$C\big(\bar\upphi; (\upvarphi, \bar\upzeta) \big)$
is a $\bar\G$ 1-cocycle. 
If $C(\bs\upsilon,\, \bs u) =\exp\{-\tfrac{i}{\hbar} c(\bs\upsilon,\, \bs u)\}$,
then $c(\bs\upsilon,\, \bs u)^{(\upvarphi,\, \bar\upzeta\,)} 
= c(\bs\upsilon,\, \bs u) +  c\big(\bar\upphi;  (\upvarphi,\, \bar\upzeta\, )  \big)$.
%\begin{align}
% c(\bs\upsilon,\, \bs u)^{(\upvarphi,\, \bar\upzeta\,)} 
%= c(\bs\upsilon,\, \bs u) +  c\big(\bar\upphi;  (\upvarphi,\, \bar\upzeta\, )  \big). 
%\end{align}
Correspondingly, the action of Lie$\bar\G=\diff(N)\,\oplus$ Lie$\bar{\mathcal G}$ is 
\begin{align}
\label{Cocyclic-dressing-2nd-kind-inf}
\delta_{(Y, \bar\upxi)}\, C(\bs\upsilon,\, \bs u)
=
C(\bs\upsilon,\, \bs u)\ a\big( (Y, \bar\upxi);\bar\upphi \big),
\end{align}
for $(Y, \bar\upxi\,)\in \bar\G$, 
and with
$a\big( (Y, \bar\upxi); \bar\upphi \big)$  a Lie$\bar\G$ 1-cocycle. 

\noindent 
The cocyclic dressing 
transforms under $\bar{\bs\G}$ and  Lie$\bar{\bs\G}$ as
\begin{equation}
\label{cocyclic-dressing-2nd-kind-trsf}
\begin{aligned}
C(\bs\upsilon, \bs u)^{(\bs\upvarphi,\, \bar{\bs\upzeta}\,)}&=
C(\bs\upsilon, \bs u)\, C(\bs\upvarphi,\, \bar{\bs\upzeta}\,),\\  
\delta_{(\bs Y, \bar{\bs\upxi})}\, c(\bs\upsilon, \bs u)
&=  i\hbar\,  a( \bs Y, \bar{\bs\upxi}).
\end{aligned}
\end{equation}
The dressed PI \eqref{Dressed-PI}-\eqref{Dressed-PI-2} thus $\bar{\bs\G}$-transforms as 
\begin{equation}
\begin{aligned}
\label{2nd-kind-dressed-PI}   
Z[\bar\upphi]^{(\bs\upvarphi, \bar{\bs\upzeta})}
  &=
C (\bs\upvarphi, \bar{\bs\upzeta}\, ) \-
  Z[\bar\upphi], \\
\exp\tfrac{i}{\hbar} \{W[\bar\upphi]^{(\bs\upvarphi, \bar{\bs\upzeta})}\}
  &=\exp\tfrac{i}{\hbar} \{
W[\bar\upphi]
+
c\big(\bs\upvarphi, \bar{\bs\upzeta}\, )  \}.
\end{aligned}
\end{equation}
Infinitesimally, this gives 
\begin{equation}
\label{2nd-kind-anomaly}
\delta_{(\bs Y, \bar{\bs\upxi})}\, W[\bar\upphi] 
= 
i\hbar\, a\big(\bs Y, \bar{\bs\upxi}\,),
\end{equation}
i.e., the \emph{anomaly for transformations of the second kind} $\bar{\bs\G}$.
This is the anomalous Ward identity for the dressed effective functional. 
The $\bar{\bs\G}$-anomaly  has the same \emph{functional} expression as the initial $\bs\G$-anomaly.

$\bullet$ In the second case,  $\bar{\bs\G}$  does act non-trivially on bare fields, 
$\upphi^{(\upvarphi,\, \bar\upzeta)}\neq\upphi$.
Then, there is no shortcut to find the $\bar\G$-transformation of dressed fields, ${\bar\upphi}^{(\upvarphi, \bar\upzeta)}$, which has to be computed on a case by case basis. 
The same goes for the $\bar\G$-transformations of the dressed PI and effective functional, which are,  
generically, 
\begin{equation}
\label{2nd-kind-dressed-PI-bis} 
Z[\bar\upphi]^{(\bs\upvarphi, \bar{\bs\upzeta})}
=
\bar C\big(\bar\upphi;  (\bs\upvarphi, \bar{\bs\upzeta} )  \big)\-\ 
  Z[\bar\upphi] 
\end{equation}
so $W[\bar\upphi]^{(\bs\upvarphi, \bar{\bs\upzeta})}
  =
W[\bar\upphi]
-
\bar c\big(\bar\upphi;  (\bs\upvarphi, \bar{\bs\upzeta} )  \big)$, 
with $\bar C=\exp \{-\tfrac{i}{\hbar}\bar c\}$  a $\bar\G$ 1-cocycle  distinct from the original $\G$ 1-cocycle $C=\exp \{-\tfrac{i}{\hbar}c\}$.
Infinitesimally, 
\begin{align}
\label{2nd-kind-anomaly-2}
\delta_{(\bs Y, \bar{\bs\upxi})}\, W[\bs\upphi] 
= 
i\hbar\, \bar a(\bs Y, \bar{\bs\upxi}),
\end{align}
giving the $\bar{\bs\G}$-anomaly, whose functional expression is a priori distinct from that of the original $\bar{\bs\G}$-anomaly.
\smallskip

Eqs. \eqref{anomalous-Z-W}-\eqref{Linear-anom-Z-W}, \eqref{Dressed-PI-2}-\eqref{cocyclic-dressing-kills-anomaly}, and \eqref{2nd-kind-anomaly}-\eqref{2nd-kind-anomaly-2} constitute the general form of the \emph{anomaly seesaw mechanism}.
Any physical information carried by the original $\bs\G$-anomaly recurs in the relational formulation via the $\bar{\bs\G}$-anomaly,
which signals a breakdown of physical frame covariance at the quantum level. 

A notable example of the above mechanism is the shift between Lorentz and Einstein anomalies --
see, e.g., \cite{Bertlmann} chap. 12 (eqs. (12.469)-(12.470) and eqs. (12.473)-(12.475)) and \cite{Bonora2023} chap. 15 (eqs. (15.49)-(15.64)): 
There the tetrad field plays the role of (field-dependent) $\hh=\mathcal{SO}(1,3)$-dressing field $\bs u=e$, with $\upphi=\omega$ the spin connection and $\bar\upphi=\Gamma$ the affine connection. 
The cocyclic dressing $C(\bs u)=C(\upphi; \bs u)=C(\omega; e)=\exp\{- \tfrac{i}{\hbar} c(\omega; e)\}$ is the \emph{Bardeen-Wess-Zumino term} implementing the Einstein-Lorentz anomaly shift.

\bibliography{Biblio16.5}

\end{document}